%% file: main.tex
\documentclass[]{spie}  %>>> use for US letter paper  .
\usepackage{amsmath,amsfonts,amssymb}
\usepackage{graphicx}
\usepackage[colorlinks=true, allcolors=blue]{hyperref}
\usepackage{xspace}
\usepackage{caption}
\usepackage{subcaption}

\input{commands.tex} 

\title{Tricouplers for nulling interferometry with photonic integrated circuits}

\author[a]{Anusha Pai Asnodkar*}
\author[a]{Nemanja Jovanovic}
\author[b]{Harry-Dean Kenchington Goldsmith}
\author[c]{Ahmed Sanny}
\author[c]{Yoo Jung Kim}
\author[d]{Hani Nejadriahi}
\author[e]{Isabelle Rivera}
\author[c]{Michael P. Fitzgerald}
\author[a]{Dimitri Mawet}
\author[c]{Pradip Gatkine}
\affil[a]{Department of Astronomy, California Institute of Technology, 1200 East California Boulevard, Pasadena, 91125, CA, USA}
\affil[b]{Astrophotonics Australia, Canberra, ACT, Australia}
\affil[c]{Physics \& Astronomy Department, University of California, Los Angeles, 475 Portola Plaza, Los Angeles, 90095, CA, USA}
\affil[d]{NASA Jet Propulsion Laboratory, 4800 Oak Grove Drive, La Cañada Flintridge, 91011, CA, USA}
\affil[e]{Department of Aerospace Engineering, University of Maryland, College Park, MD 20742, USA}

\authorinfo{Further author information: (Send correspondence to A.P.A)\\
\footnote{*}A.P.A.: E-mail: apaiasno@caltech.edu}

\begin{document} 
\maketitle

\begin{abstract}
Solar System analog gas giants and habitable-zone terrestrial planets are observationally elusive to conventional exoplanet detection and characterization techniques, i.e. transits, radial velocities, and direct imaging. Long baseline nulling interferometry across multiple apertures suppresses starlight and enables detection of faint planetary signals at higher spatial resolution than traditional coronagraphs on single-aperture telescopes. Leveraging technological advancements from the telecommunications industry, photonic integrated circuits (PICs) offer a promising platform for performing the optical operations necessary for astronomical applications, including phasing and beam combination for nulling interferometry in both long-baseline and cross-aperture configurations. PICs provide compact, scalable architectures with reduced sensitivity to alignment as well as thermal and mechanical perturbations compared to bulk optics. However, their design and manufacturing precision remain insufficient for the stringent requirements of exoplanet instrumentation. Here, we investigate the nulling capabilities of photonic tricouplers, devices composed of three equal-width waveguides that are geometrically predisposed to produce achromatic nulls upon beam combination through their symmetric construction. In the laboratory, we characterize null depths in monochromatic light at 1.55 \micron with devices on a planar silica-on-silicon platform. In broadband $H$-band light, we explore chromatic effects from components such as thermo-optic phase modulators used for fine-phasing. Our ongoing efforts towards maturation of PICs for direct detection and atmospheric characterization of exoplanets will support scalable testing and deployment of high-contrast technologies for future space-based observatories, such as the Habitable Worlds Observatory. 
\end{abstract}

\keywords{photonics, integrated circuits, nulling interferometry, high contrast, exoplanet}

\section{INTRODUCTION}
\label{sec:intro}  % \label{} allows reference to this section
Exoplanets near primordial snow lines occupy a critical regime for understanding planet formation and habitability. Demographics spanning a broad range of planet-star separations combining direct imaging, radial velocity, and microlensing surveys suggest that giant planet occurrence rates peak near the water ice line around $\sim$3 AU \cite{Fulton2021}. The water inventory of terrestrial planets, a key factor in determining their potential habitability, is shaped by both in situ accretion during formation and the subsequent delivery of volatiles by icy planetesimals. These processes depend strongly on the planet's formation relative to the snow line and on dynamical interactions with nearby giant planets \cite{Morbidelli2000}. Despite the wealth of scientific insight offered by this population, planets in this regime remain observationally elusive. Their relatively large orbital separations result in low transit probabilities and weak radial velocity signals. Moreover, their small angular separations from their host stars often fall below the inner working angles (IWAs) of traditional coronagraphs ($\sim$2\textlambda/D).

Long-baseline nulling interferometry enables direct detection of exoplanets by suppressing stellar light at high spatial resolution, thereby facilitating studies of exoplanet demographics, orbital dynamics, and atmospheric composition through spectroscopic characterization. Presently, nulling interferometers have not demonstrated the contrast performance necessary to detect Solar System analog giants or habitable-zone terrestrial planets. Achieving such contrasts requires exquisite control of beam combination and optical path length. These requirements become increasingly demanding for bulk-optic architectures that are sensitive to alignment errors, thermal drifts, and mechanical perturbations. 

Photonic integrated circuits (PICs) are a promising platform for performing the optical operations needed for interferometry, i.e. beam combining and phasing \cite{Jovanovic2023}. Photonics allow for full electric field control as light propagates within single and multimode waveguides. The miniaturized footprint of PICs facilitates scalable fabrication and finer control than standard optical fibers. Moreover, the integrated platform minimizes wavefront errors from thermal and mechanical perturbations. Leveraging technological advances from the telecommunications industry, the field of astrophotonics has progressed towards realizing integrated beam combiners for nulling interferometry, culminating in successful on-sky demonstrations such as the GLINT instrument at the Subaru Telescope and the Asgard/NOTT instrument at the Very Large Telescope Interferometer \cite{Norris2020, Spalding2024, Laugier2023, Sanny2026}.

In this work, we explore the chromatic nulling performance of lithographic tricouplers with thermo-optic phase modulators (TOPMs) for beam-combination and two-beam interferometry. Our laboratory demonstration extends previous efforts \cite{Hsiao2010,Martinod2021b,Jovanovic2025} by incorporating Mach-Zender interferometers (MZIs) for intensity-matching and evaluating chromatic null degradation. This work is a direct continuation of Jovanovic \textit{et al.}\cite{Jovanovic2025}.

\section{METHODS}

\subsection{Testbed}

\begin{figure}[t]
    \centering
    \includegraphics[width=0.75\linewidth]{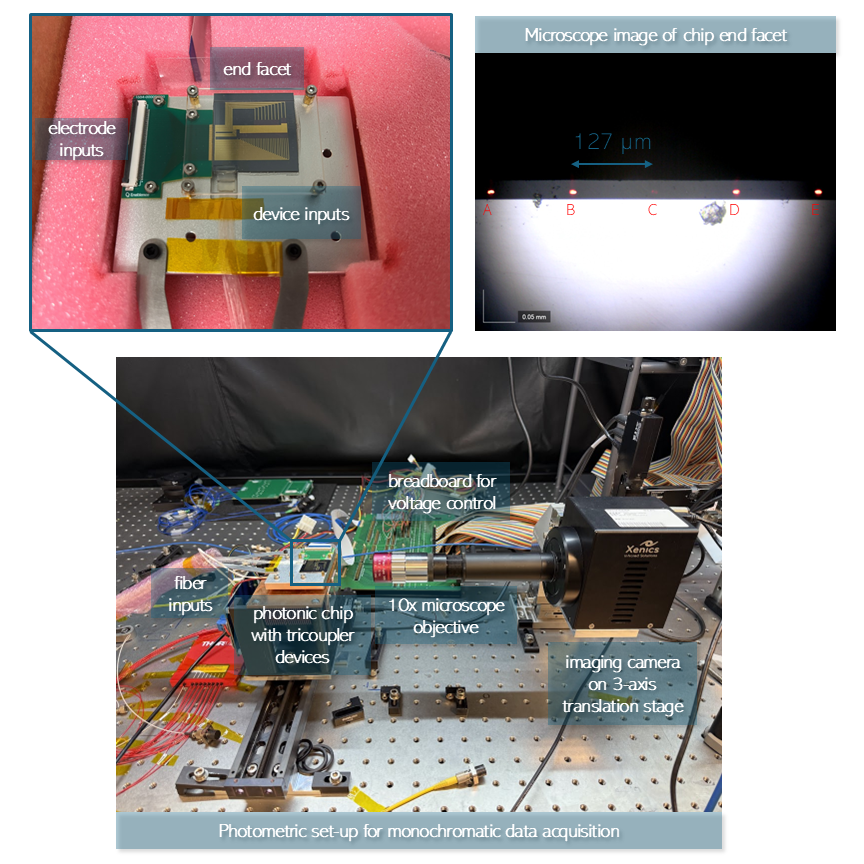}
    \caption{Photos of chip (top left), testbed (bottom), and microscope imaging of chip end facet with output waveguide cores illuminated at 635 nm (top right)}
    \label{fig:testbed}
\end{figure}

The PICs tested in this work were lithographically fabricated by Enablence on a silica-on-silicon platform. The device consists of a SiO$_2$ cladding layer deposited on a silicon substrate, with waveguide cores composed of Ge-doped SiO$_2$ providing a 2\% index contrast (see Figure \ref{fig:testbed}). The input facet of the chip is coupled to a v-groove array of polarization-maintaining fibers, while the output facet remains bare to facilitate both imaging (setup with microscope objective and translatable detector shown in Figure \ref{fig:testbed}) and fiber coupling of the emitted light. 

Electrodes for active thermo-optic modulation were lithographically deposited on the chip and wire-bonded to the outer packaging. A multichannel ribbon cable interface allows for external electronic control on a breadboard. To introduce a phase shift, a voltage is applied to a TOPM, which locally heats the region of waveguide in contact with and surrounding the electrode. Since the waveguide refractive index is a function of temperature and wavelength, i.e. $n_\mathrm{core} = n_\mathrm{core}(\lambda, T)$, this heating consequently induces an optical path difference relative to the inactive state. Voltage control of the TOPMs is programmatically operated by an Nicslab XPOW-120AX-CCvCV-U multichannel voltage controller interfaced with a Nicslab M2 breadboard. For temperature stabilization, the chip is mounted on an aluminum block, which is coupled to a thermoelectric cooler (TEC) sandwiched between the aluminum block and a copper heatsink with a fan radiator. The TEC is driven by an Arduino-based PID feedback controller using temperature readout from a thermistor sensor secured to the bottom of the aluminum block, directly underneath the PIC, with thermal paste. 

We acquire data in two different modes: photometric monitoring of multiple monochromatic outputs and spectral monitoring of a single broadband output. We take images for photometry using a Xenics XEVA-2.5-320 camera in monochromatic light with a Thorlabs TLX2 laser input at 1590 nm, which allows us to measure all outputs of a given device simultaneously (see Figure \ref{fig:ImageExamples}) and also explore polarization effects. The power of the laser input beam is modulated with the laser's internal variable optical attenuator (VOA) as well as an external VOA (Thorlabs V1550PA) to capture the dynamic range of the chip output modulation while maintaining operation within the camera's linear response regime. For polarization-filtered measurements, we install a polarizer at the backend of the chip, right before the imaging optics and detector. Since these measurements typically yielded very deep nulls, we also mount a pinhole-like mask between the end facet of the chip and the polarizer. This mask selectively transmits light from only one output, the nulling port, to minimize contamination from scattering and diffraction of the bright outputs that artificially degrade null depth measurements by contributing background to the extracted photometry. In a separate data acquisition mode, we also obtain polychromatic data with a Thorlabs OSA202C optical spectrum analyzer (OSA) using a Thorlabs S5FC1550P-A2 superluminous diode broadband source around 1550 nm. This requires coupling a single output into a single mode fiber to route to the OSA, so this data acquisition mode cannot be used to monitor all outputs simultaneously.

\subsection{Photonic components and circuits}
\begin{figure}[t]
    \centering
    \begin{subfigure}[b]{0.2\linewidth}
        \centering
        \includegraphics[width=\linewidth]{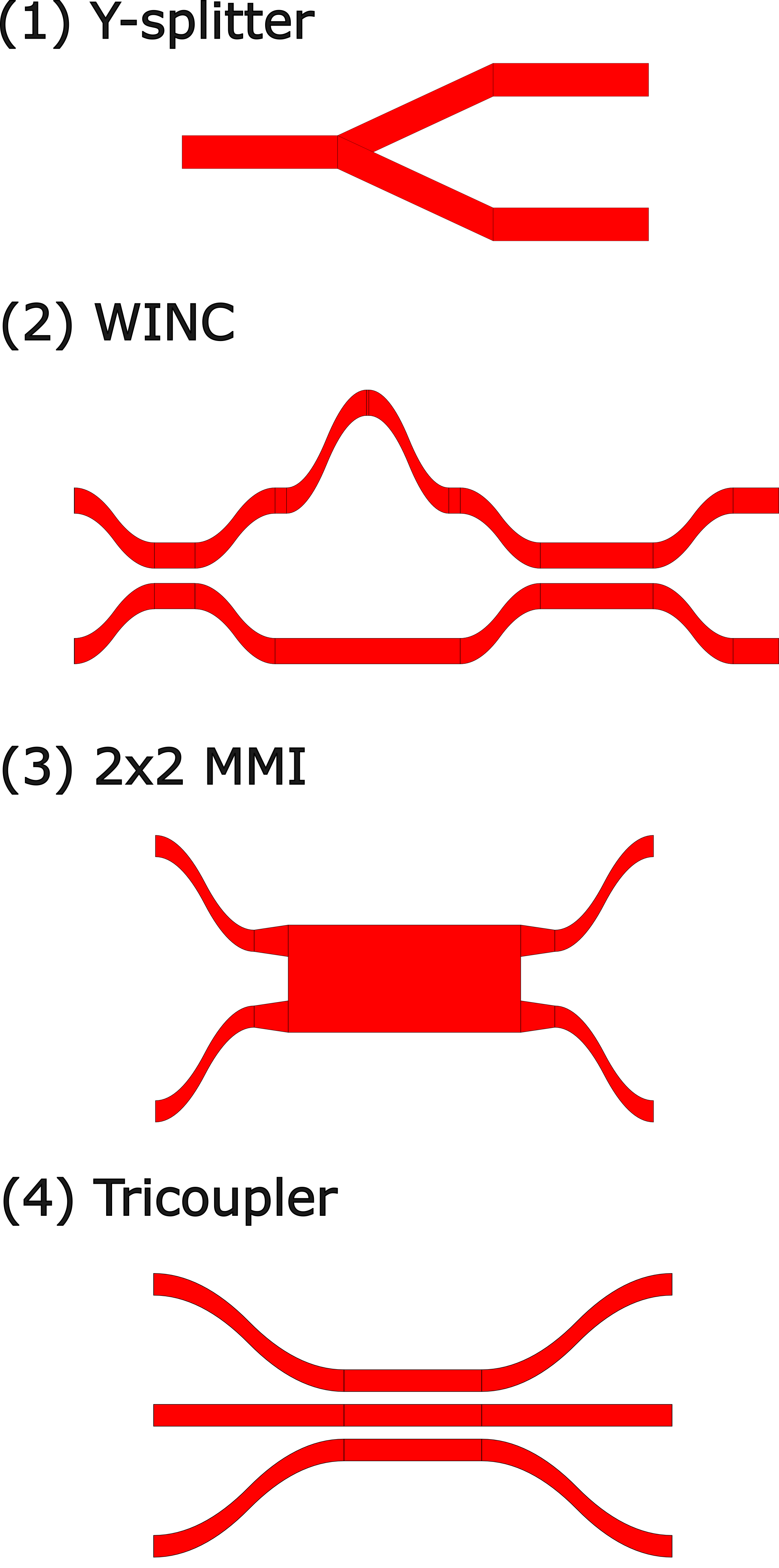}
        \caption{}
        \label{fig:PassiveComponents}
    \end{subfigure}
    \hspace{1em}
    \begin{subfigure}[b]{0.41\linewidth}
        \centering
        \includegraphics[width=\linewidth]{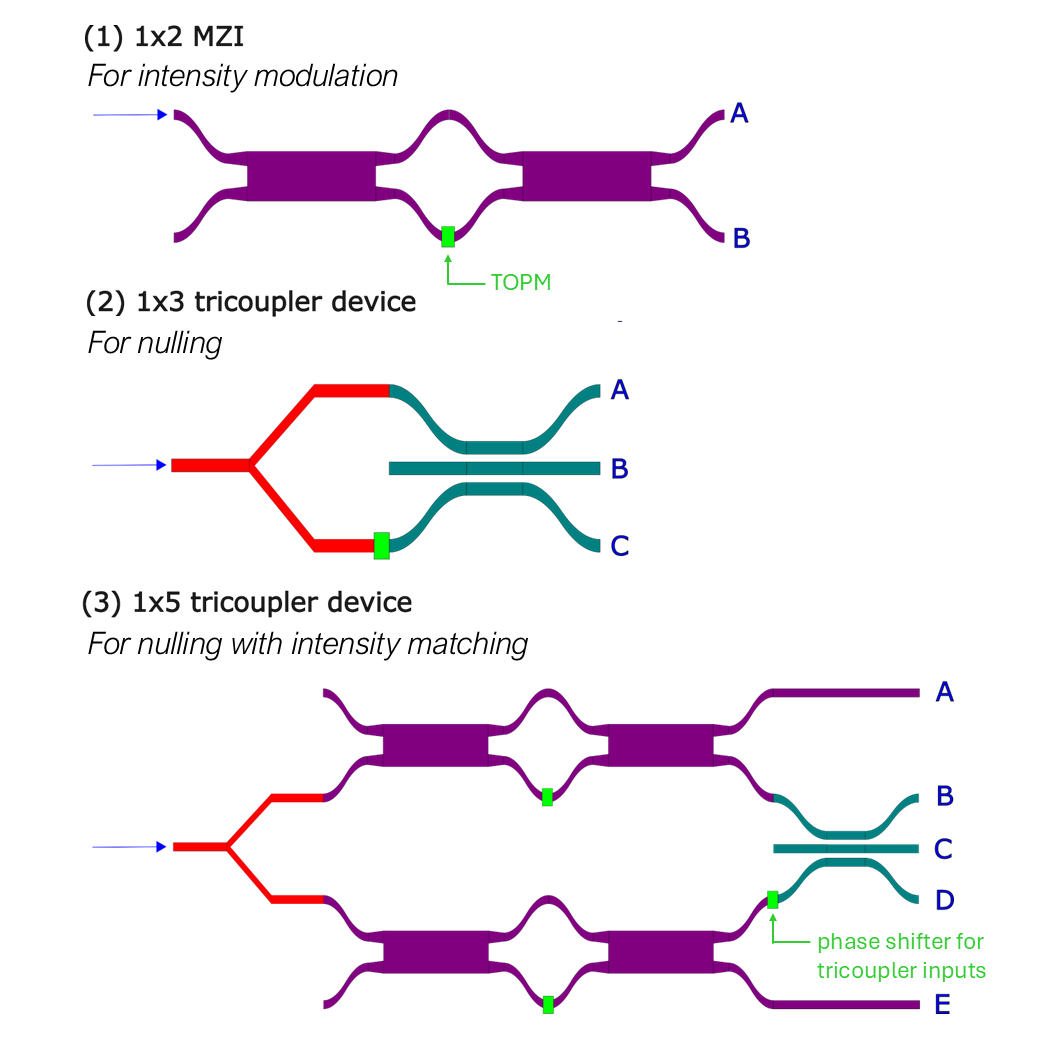}
        \caption{}
        \label{fig:ActiveDevices}
    \end{subfigure}
    \caption{Schematics of: (a) passive components and (b) active devices. The passive components presented are used for 50:50 splitting (Y-branch splitter and WINC), building Mach-Zehnder interferometers (2x2 MMI), and beam combination (tricoupler) in the subsequent active devices. The active devices feature 1x2 MZIs (highlighted in purple) for intensity matching and tricouplers (highlighted in teal) for beam combination. TOPMs are indicated with green boxes.}
    \label{fig:DeviceSchematics}
\end{figure}

\begin{figure}[b]
    \centering
    \includegraphics[width=0.5\linewidth]{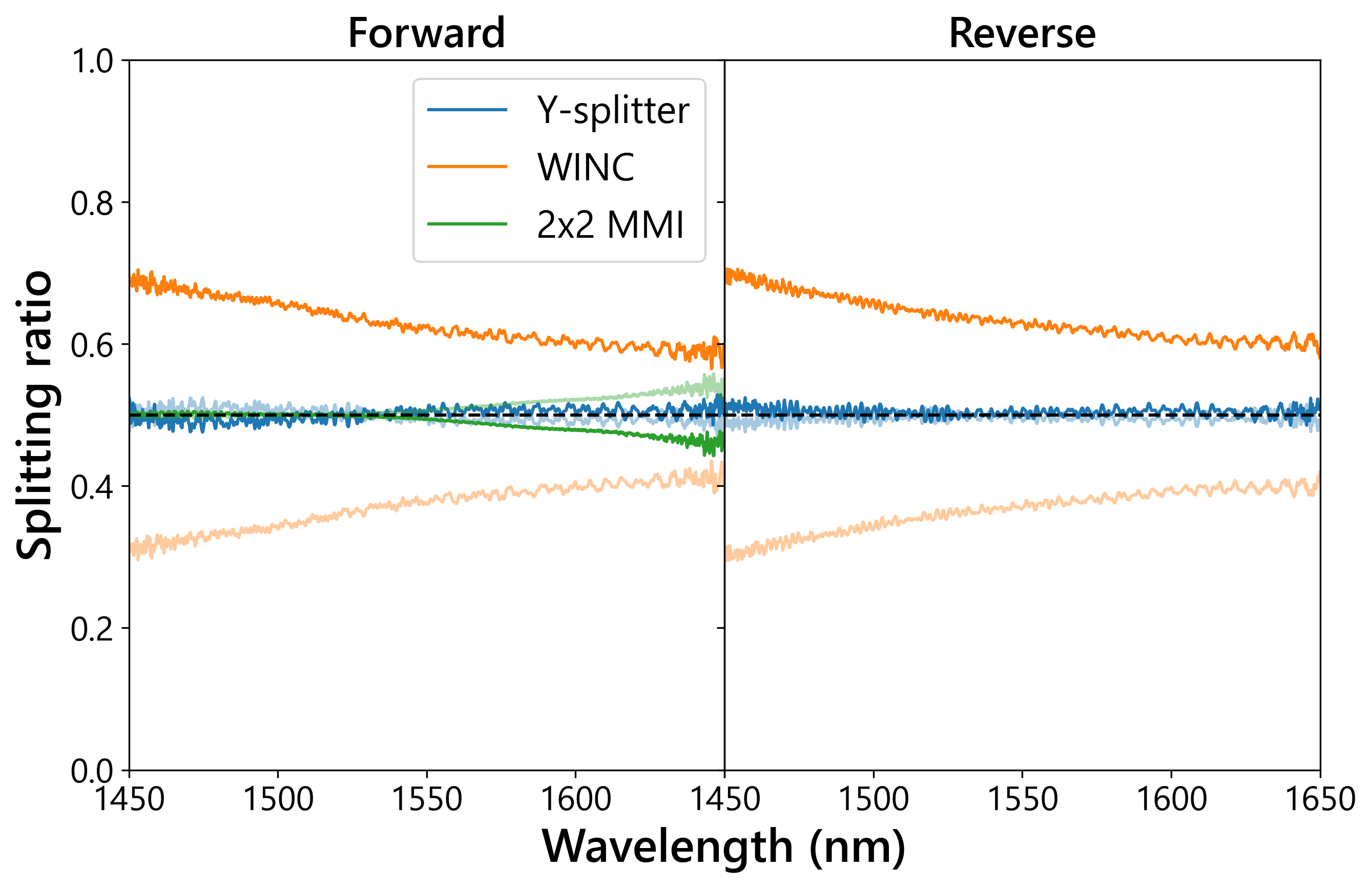}
    \caption{Polychromatic splitting ratios of a Y-splitter, single-injection WINC, and single-injection 2x2 MMI. Solid and semi-transparent curves represent the two outputs of each device. Left panel presents splitting ratios injecting light into device in forwards direction through the chip input fibers, while the right panel presents splitting ratios from reverse injection from the end facet of the chip (note that we do not have reverse injection data for the 2x2 MMI). These passive components should behave the same way in forward and reverse, so chromatic differences point to differences in input }
    \label{fig:SplitterSplitting}
\end{figure}

Our initial attempts with two-input injection exhibit significant instability due to fluctuating path-length differences between the inputs, highlighting the need for more stringent testbed environmental control. As a result, we focus our analysis in this work on single-input devices that generate the two coherent inputs on-chip using integrated Y-splitters. Since nulling interferometry requires two antisymmetric inputs of equal intensity, we begin by characterizing the passive chromaticity of the splitting components used in these devices. In this section, we evaluate the chromaticity of 3 different candidate components for broadband splitting (see Figure \ref{fig:PassiveComponents}): (1) Y-splitters, (2) wavelength independent couplers (WINCs), and (3) 2x2 multimode interferometers (MMIs). 

For this first demonstration, we simplify our experimental setup by splitting a single beam injection on-chip to simulate the two beams coming from two telescope apertures/sub-apertures for combination in the manner of a traditional Bracewell nulling interferometer \cite{Bracewell1978}. Therefore, the splitter must produce 50:50 splitting across the entire bandpass of interest. A Y-splitter begins with a single input and symmetrically branches off into two separate arms. The ideal transformation of the electric field performed by a 50:50 Y-splitter can be described by the transfer matrix:
\[
\mathbf{M}_{\mathrm{Ysplitter}}
=
\frac{1}{\sqrt{2}}
\begin{bmatrix}
1 \\
1
\end{bmatrix}
\]

\begin{figure}[t]
    \centering
    \includegraphics[width=0.8\linewidth]{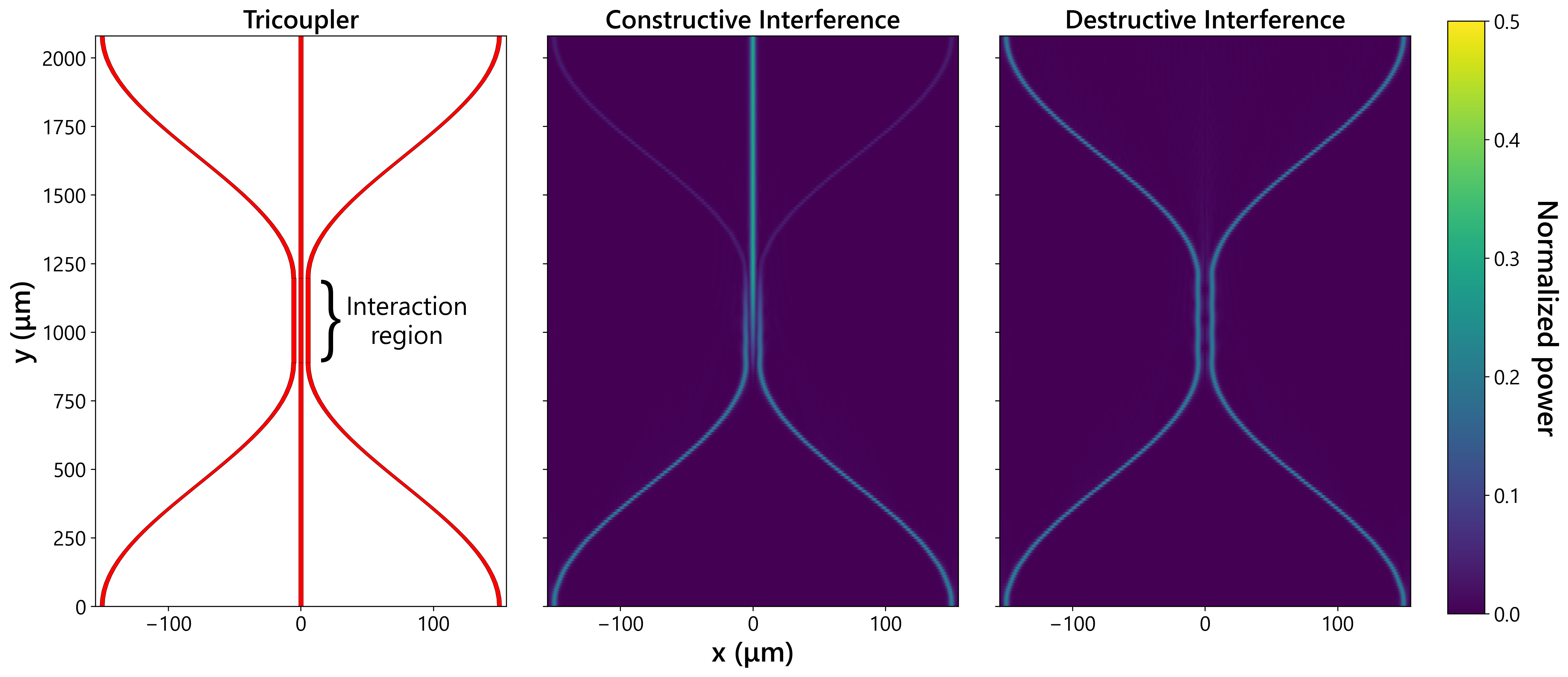}
    \caption{Tricoupler waveguide structure (left) along with simulations of beam propagation through the device in the cases of constructive (middle) and destructive interference (right). Tricouplers serve as beam combiners for inputs injected in the left and right waveguides through evanescent coupling in the interaction region. When a \textpi phase shift is imposed between the two outer inputs where light from two apertures is injected, the on-axis starlight is achromatically nulled in the central port (destructive interference). Meanwhile, off-axis planet light will have different path lengths to the two telescope apertures, and can still constructively interfere in the central port for most sub-\textlambda/D planet-star separations.}
    \label{fig:Tricoupler_beamProp}
\end{figure}

\begin{figure}[b]
    \centering
    \includegraphics[width=0.8\linewidth]{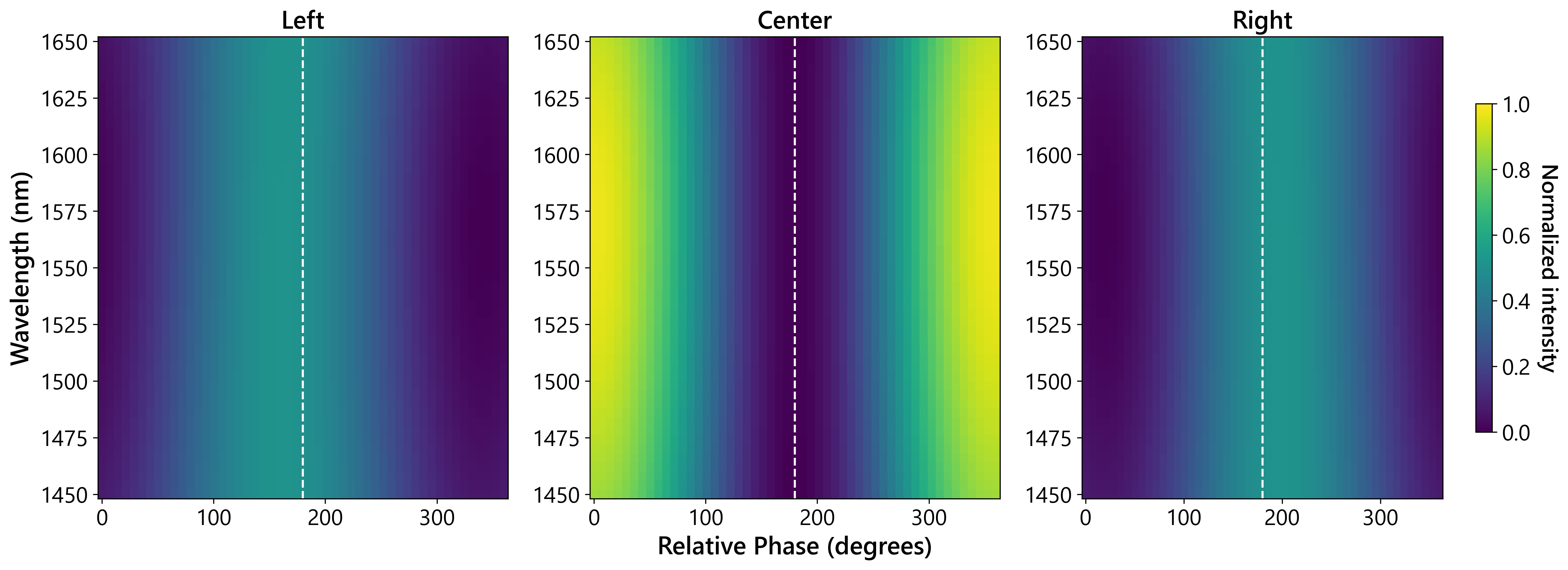}
    \caption{Splitting ratios as a function of input phase difference and wavelength simulated for an ideal tricoupler.}
    \label{fig:Tricoupler_outputs}
\end{figure}

In our experiment, a WINC is an asymmetric two-waveguide directional coupler, with one waveguide mapping out a different path than the other to enable a flat coupling response across a broad bandpass. While this device has two inputs and two outputs, for our purposes we only inject light into one input to use it as a splitter. Alternatively, a 2×2 MMI accepts two single-mode inputs, combines them in a multimode slab region where the optical fields interfere, and then couples the resulting fields into two single-mode output waveguides. Again, we only inject into one of the inputs to use this component as a splitter. The transfer matrix of both a WINC and 50:50 splitting 2x2 MMI is:
\[
\mathbf{M}_{\mathrm{WINC}} = \mathbf{M}_{\mathrm{MMI}}
=
\frac{1}{\sqrt{2}}
\begin{bmatrix}
1 & i\\
i & 1
\end{bmatrix}
\]

In Figure \ref{fig:SplitterSplitting}, we compare the splitting performance of the Y-splitter, WINC, and 2x2 MMI. The Y-splitter and WINC are test components provided on our active chip, while the 2x2 MMI probed in this section is on our passive chip. Figure \ref{fig:SplitterSplitting} demonstrates the most achromatic 50:50 splitting from the Y-splitter, although the 2x2 MMI offers a close alternative at the bluer end of the bandpass.

The tricoupler is a symmetric three-waveguide directional coupler. For two-beam nulling interferometry, the tricoupler can be used as a beam combiner by injecting coherent light from two apertures in the two outer ports (see Figure \ref{fig:Tricoupler_beamProp}). In the interaction region of the tricoupler, the waveguides are brought close together and the light propagating along the outer ports evanescently couples and interferes across all three waveguides. If the two input beams are offset in phase by $\pi$, the two beams destructively interfere to produce an achromatic null that persists along the central port (see Figure \ref{fig:Tricoupler_outputs}). In this configuration with starlight on-axis, the throughput of off-axis light, e.g. from an exoplanet, in the central port is maximized at an angular separation of 0.5 $\lambda$/D. In our experiments, all devices are lithographically fabricated, so our tricouplers are planar in geometry (as opposed to the 3D ULI tricouplers extensively explored in the literature \cite{Martinod2021a, Klinner-Teo2022, Spalding2024}), and can be described by a transfer matrix of the form:
\[
\mathbf{M_\mathrm{tricoupler}} =
\begin{bmatrix}
T_1 e^{i\psi_{T_1}} & K_1 e^{i\psi_{K1}} & K_2 e^{i\psi_{K_2}} \\
K_1 e^{i\psi_{K_1}} & T_2 e^{i\psi_{T_2}} & K_1 e^{i\psi_{K_1}} \\
K_2 e^{i\psi_{K_2}} & K_1 e^{i\psi_{K_1}} & T_1 e^{i\psi_{T_1}}
\end{bmatrix}
\]
Upon imposing the unitary condition, the planar tricoupler adheres to the following constraints:
\begin{itemize}
    \item $K_1^2 + K_2^2 + T_1^2 = 1$
    \item $2 K_1^2 + T_2^2 = 1$
    \item $K_1^2 + 2 K_2 T_1 \cos(\psi_{K_2} - \psi_{T_1}) = 0$
\end{itemize}
Here $T_i$ represents a transmission coefficient and $K_i$ represents a coupling coefficient. Notably, $\Delta\psi = \psi_{K_2} - \psi_{T_1}$ is the phase difference between the tricoupler outputs. For a 3D triangular tricoupler, this phase difference is typically set to  $\Delta\psi = \frac{2 \pi}{3}$, which arises when the device is designed to have equal transmission and coupling coefficients. For a planar tricoupler, the phase difference depends on $K_1$, $K_2$, and $T_1$ as given by the third constraint. 

While the null of on-axis stellar light in the central port of a planar tricoupler is achromatic by design, the throughput of off-axis planet light in the central port is chromatic. For a discussion of tapered tricoupler designs to optimize exoplanet throughput and fringe tracking with the outer bright ports, see Kenchington Goldsmith \textit{et al.}\cite{Goldsmith2026}. In this work, we exclusively present the experimental validation of planar tricouplers with uniform waveguide widths.

\section{RESULTS}

\subsection{Chromaticity of MMI-based intensity modulators}
\label{sec:MZI}
 
\begin{figure}[t]
    \centering
    \includegraphics[width=0.6\linewidth]{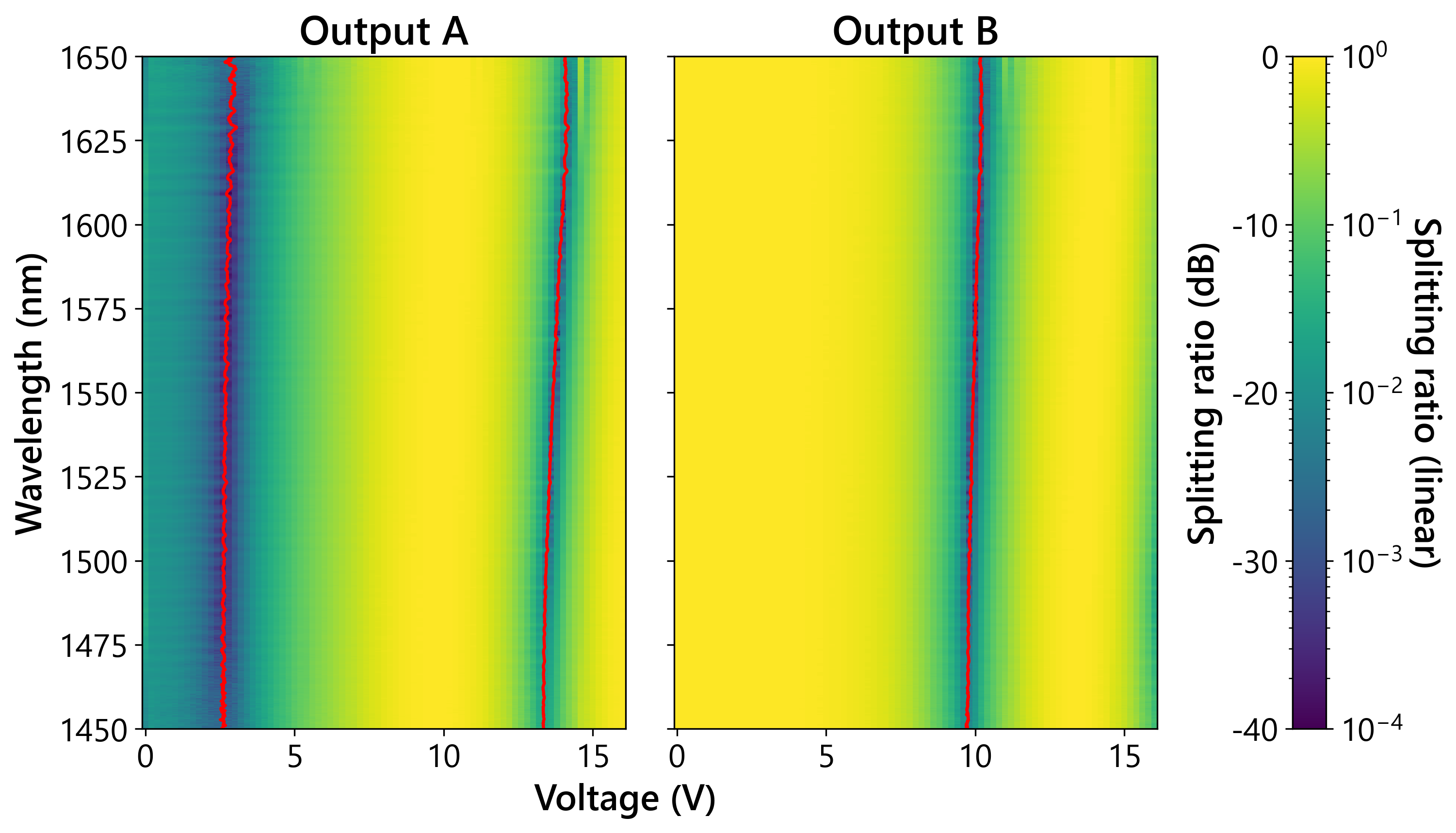}
    \caption{Wavelength-dependent splitting ratios of the 1x2 MZI device as the TOPM is modulated. Sloped nulls (red traces) indicate chromatically varying phase shifts.}
    \label{fig:1x2MZI}
\end{figure}

The 1x2 MZI in Figure \ref{fig:ActiveDevices} is adopted in some of our interferometric circuits as an intensity modulator on both arms for intensity matching before beam combination (see the 1x5 tricoupler device in Figure \ref{fig:ActiveDevices}). In Figure \ref{fig:1x2MZI}, we present the chromatic splitting ratio as the voltage of the TOPM is modulated. The minima and maxima at each wavelength in this plot can be used to anchor the wavelength-dependent phase induced by the TOPM at different voltage settings. The sloped dark stripes indicate that the minimum achieved by a given output is a splitting ratio around -30 dB ($\sim$10$^{-3}$), and this minimum occurs at different voltages for different wavelengths across the bandpass of interest. While splitting ratio of an individual MMI is fairly achromatic as presented in Figure \ref{fig:SplitterSplitting}, the chromaticity of the 1x2 MZI comes from the TOPMs. This is discussed further with quantitative rigor in  \S \ref{sec:polychromatic}. Phase modulation using TOPMs is inherently chromatic because it relies on the thermo-optic modification of the waveguide refractive index, which is itself wavelength-dependent. We use this result for the 1x2 MZI to contextualize the chromaticity of the tricoupler nulling circuits in \S \ref{sec:polychromatic}.

\subsection{Nulling with tricoupler devices}
We present results for 3 tricoupler devices: 1) a 1x3 tricoupler device (2nd circuit in Figure \ref{fig:ActiveDevices}), 2) a 1x5 tricoupler device (3rd circuit in Figure \ref{fig:ActiveDevices}), and 3) a 1x5 tricoupler device with a 76$^\circ$ waveguide crossing (3rd circuit in Figure \ref{fig:ActiveDevices} with the top and bottom inputs to the tricoupler crossed). The device with a waveguide crossing is included to evaluate how crossings affect the achievable null depth of the tricoupler, as future spectrally multiplexed lithographic nulling circuits may require waveguide crossings. The 1x3 tricoupler device has one TOPM for inducing a \textpi phase shift between the two inputs to the tricoupler. The 1x5 tricoupler devices have 3 TOPMs, two for the 1x2 MZI intensity modulators to match the intensity in both arms before interference, and one for the \textpi phase shift between the two arms before feeding into the tricoupler for beam combination. We adopt a brute-force grid search as well as a more algorithmic Nelder-Mead minimization to search for the intensity modulator settings that yield the deepest nulls in monochromatic light, while still directing a small fraction ($\sim$5\%) of light to the outer dump ports, which can be used on-sky for wavefront sensing. 

\subsubsection{Monochromatic nulling}
\label{sec:monochromatic}

\begin{figure}[b]
    \centering
    \begin{subfigure}{0.5\linewidth}
        \centering
        \includegraphics[width=\linewidth]{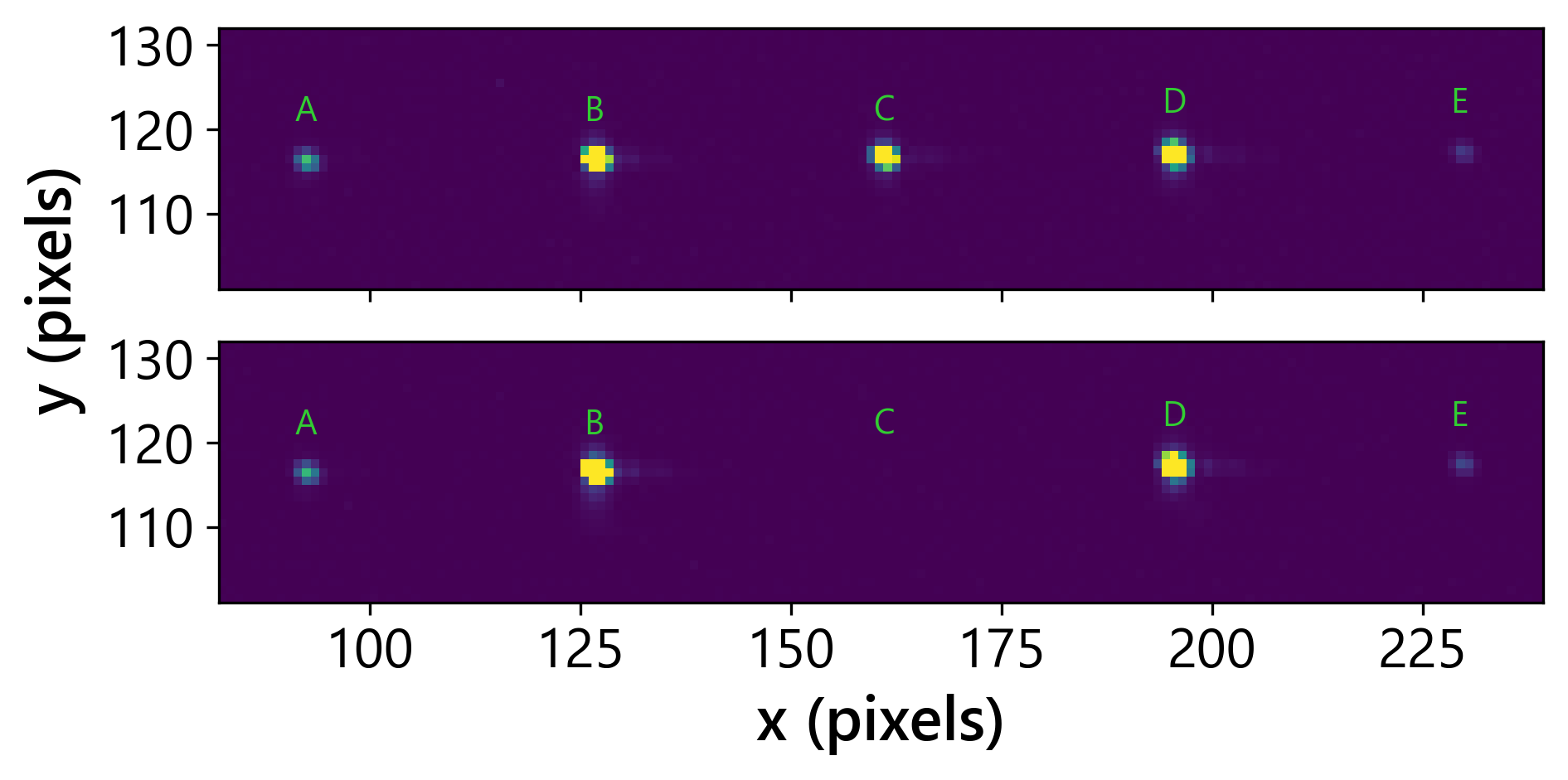}
        \caption{}
        \label{fig:ImageExamples}
    \end{subfigure}
    \hspace{1em}
    \begin{subfigure}{0.4\linewidth}
        \centering
        \includegraphics[width=\linewidth]{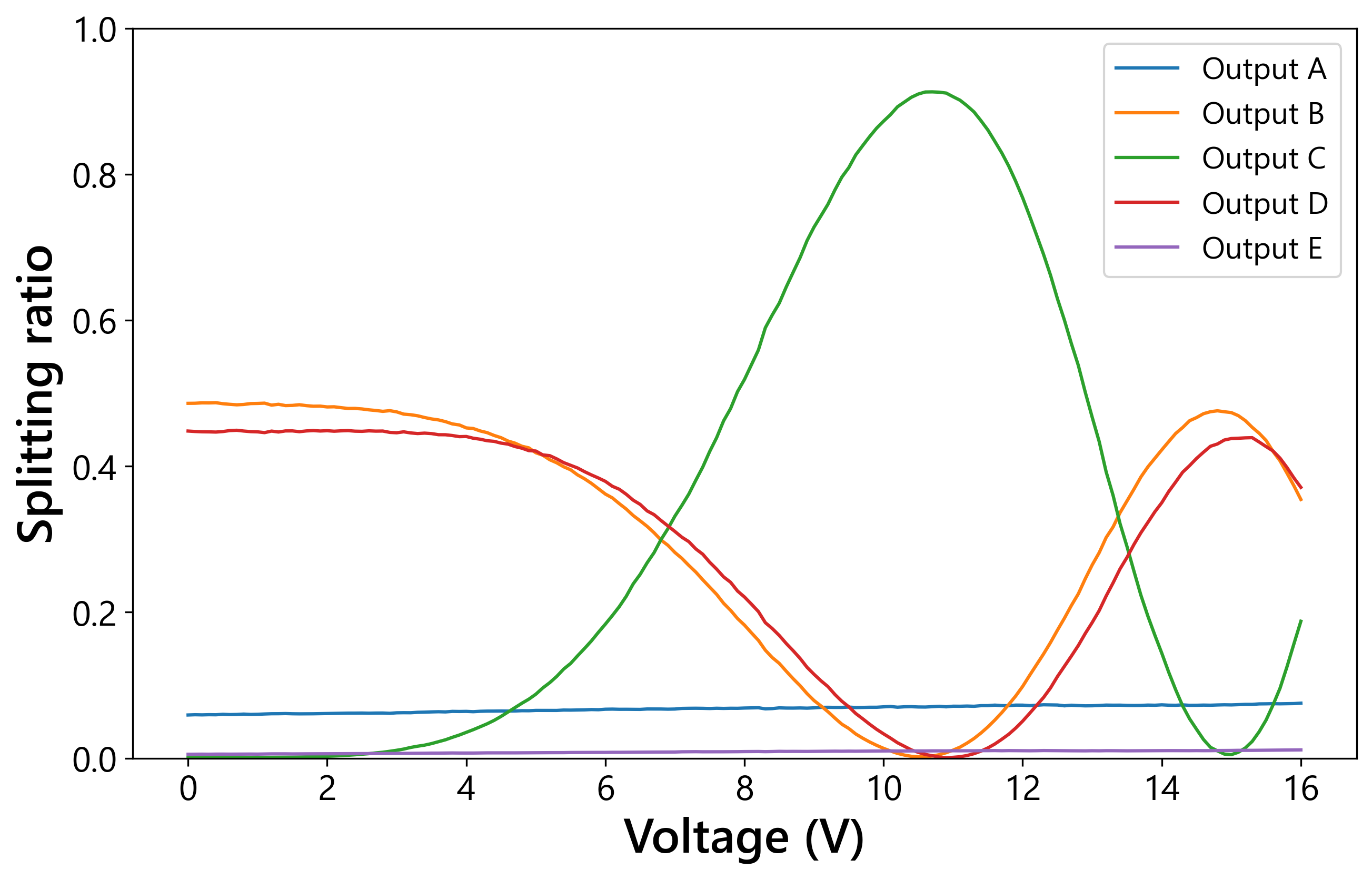}
        \caption{}
        \label{fig:PhotometryExamples}
    \end{subfigure}
    \caption{(a) Image of 1x5 tricoupler device outputs for arbitrary voltage settings showing light distributed across all 5 outputs (top) compared with voltages configured close to a minimum (null) in output C (bottom). (b) Splitting ratio of all outputs of the 1x5 tricoupler device as the phase shifter voltage is modulated.}
    \label{fig:PhotometryModulation}
\end{figure}

In monochromatic light (1590 nm), we extract photometry from all outputs for each of the three tricoupler devices while modulating the TOPM corresponding to the tricoupler input phase shifter. For the 1x5 devices, we first identify and set optimal voltage settings for the intensity modulator TOPMs before scanning the full range of the tricoupler input phase shifter. Example images of the outputs are presented in Figure \ref{fig:PhotometryModulation}, along with output intensity modulation over the full phase shifter voltage scan for all 5 outputs of the 1x5 tricoupler device.

\begin{figure}[t]
    \centering
    \begin{subfigure}{0.45\linewidth}
        \centering
        \includegraphics[width=\linewidth]{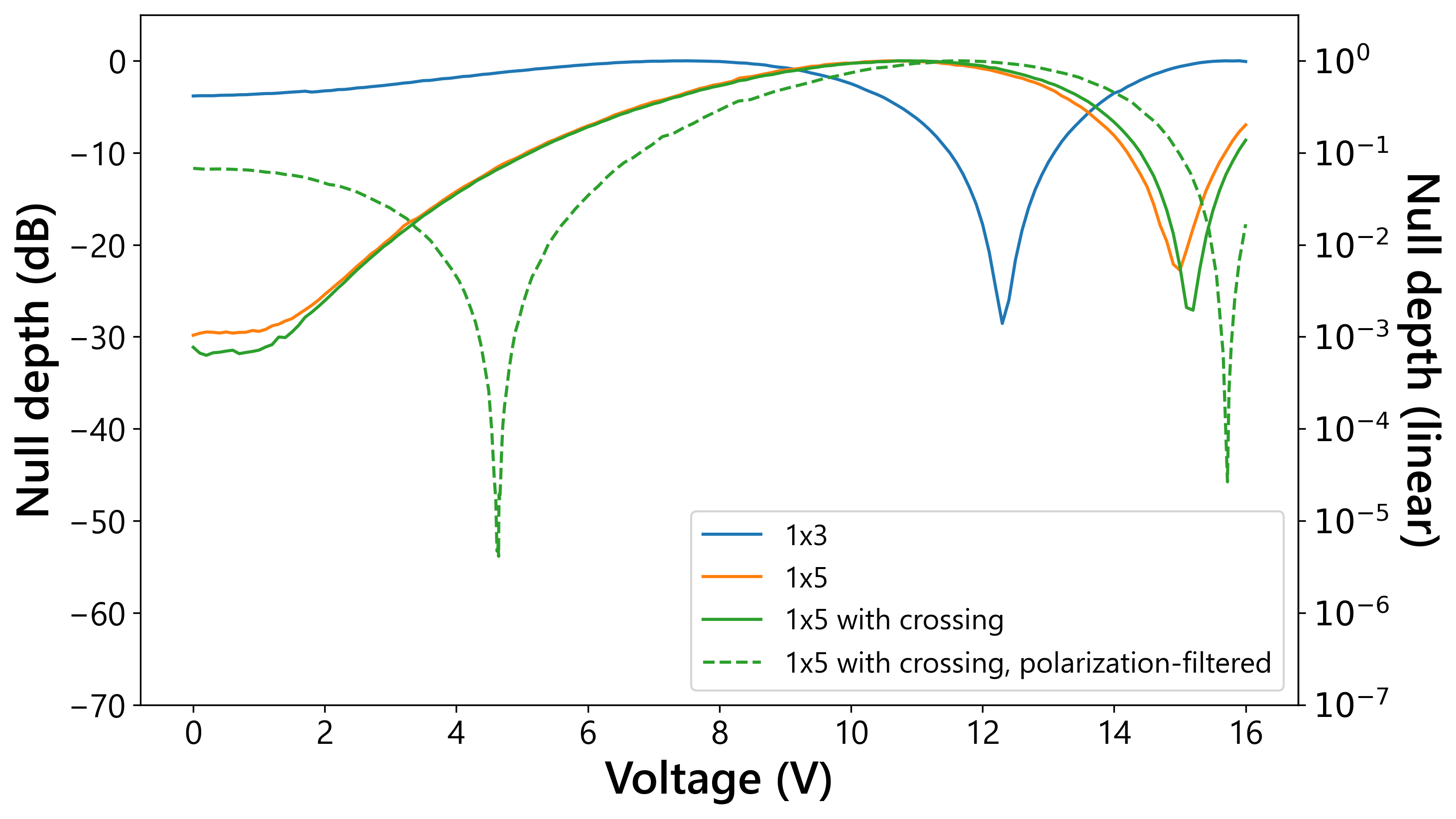}
        \caption{}
        \label{fig:NullDepthMonoAll}
    \end{subfigure}
    \hspace{1em}
    \begin{subfigure}{0.45\linewidth}
        \centering
        \includegraphics[width=\linewidth]{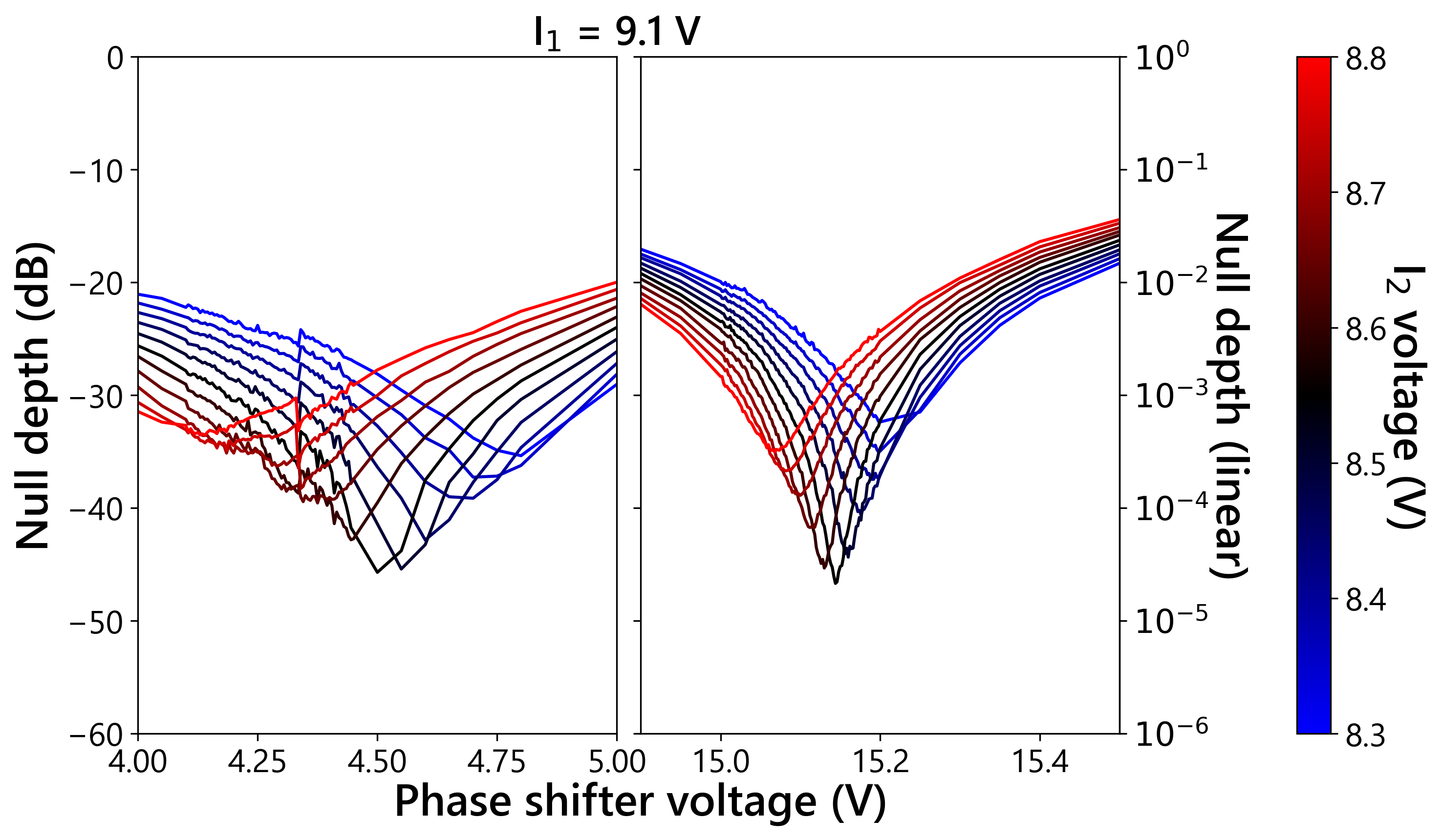}
        \caption{}
        \label{fig:NullDepthSensitivity}
    \end{subfigure}
    \caption{(a) Null depth in tricoupler central port as input phase shifter voltage is modulated for all three tricoupler devices. (b) Sensitivity of null depth to intensity modulator tuning for the 1x5 tricoupler with crossing and polarization filtering.}
    \label{fig:NullDepthMono}
\end{figure}

To measure null depth, we adopt the convention: 
\begin{equation}
    N = \frac{I_{\mathrm{min}}}{I_\mathrm{max}}
\end{equation}
where $N$ is null depth, $I_{\mathrm{min}}$ is the minimum intensity output from the nulling port (the central port of the tricoupler, i.e. output B for the 1x3 tricoupler device and output C for the 1x5 tricoupler devices), and $I_{\mathrm{max}}$ is the maximum intensity output from the nulling port. We present null depth scans for all three devices in Figure \ref{fig:NullDepthMonoAll}. Notably, we also present a polarization-filtered measurement for the 1x5 tricoupler with a crossing, represented by the green dashed curve. 

As demonstrated in Figure \ref{fig:NullDepthMonoAll}, the 1x3 tricoupler device achieves a null depth (-28.6 dB, 1.4 $\times$ 10$^{-3}$) that is comparable to the 1x5 devices, both without (-29.8 dB, 1.1 $\times$ 10$^{-3}$) and with (-32.0 dB, 6.3 $\times$ 10$^{-4}$) a crossing. This implies that an intensity imbalance may not be the limiting factor in reaching deeper null depths for these tricoupler devices. Moreover, a crossing does not introduce scattered light or phase imperfections that significantly degrade the null at these levels. Most saliently, the 1x5 tricoupler with the crossing displays an improvement in null depth by more than 2 orders of magnitude with polarization-filtering (-53.9 dB, 4.1 $\times$ 10$^{-6}$). This suggests that these devices exhibit greater birefringence than anticipated, causing the measured unfiltered null depth to be limited by the combination of two polarization states with different null conditions. Figure \ref{fig:NullDepthSensitivity} presents the sensitivity of this deep null to intensity imbalances by holding I$_1$ fixed and detuning I$_2$. For the deepest polarization-filtered nulls, null depth is highly sensitive to the intensity modulator settings, degrading by 1--2 OoM when one of the intensity modulators deviates from the optimal setting by 0.2 V.

An interesting feature of Figure \ref{fig:NullDepthMonoAll} is the shift in the null voltage between the 1x3 and 1x5 devices. This shift arises from the intrinsic phase relationships between the outputs of the MMIs used in the intensity modulators for the 1x5 devices. Consider the transformation applied to the input beam up until the tricoupler in the 1x3 tricoupler device depicted in Figure~\ref{fig:ActiveDevices}:

\begin{align*}
&\begin{bmatrix}
1 & 0 \\
0 & e^{i\Delta\phi}
\end{bmatrix}
M_{\mathrm{Ysplitter}} \\
=
& \begin{bmatrix}
1 & 0 \\
0 & e^{i\Delta\phi}
\end{bmatrix}
\frac{1}{\sqrt{2}} \begin{bmatrix}
1 \\
1
\end{bmatrix} \\
= 
& \frac{1}{\sqrt{2}}\begin{bmatrix}
1 \\
e^{i\Delta\phi}
\end{bmatrix}
\end{align*} 
Thus, to achieve a \textpi phase shift between the two inputs to the tricoupler for nulling, the phase shifter has to apply a voltage corresponding to a phase offset of $\Delta\phi = \pi$. 

For the 1x5 tricoupler device, the inital Y-splitter similarly splits the input beam into two arms of amplitude $\frac{1}{\sqrt{2}}$. The subsequent transformation of the top arm up to the tricoupler is:
\begin{align*}
&M_{\mathrm{MMI}}
\begin{bmatrix}
1 & 0 \\
0 & e^{i\Delta\psi_{\mathrm{top}}}
\end{bmatrix}
M_{\mathrm{MMI}}
\begin{bmatrix}
0 \\
\frac{1}{\sqrt{2}}
\end{bmatrix} \\
=
& \frac{1}{\sqrt{2}} \begin{bmatrix}
1 & i\\
i & 1
\end{bmatrix}
\begin{bmatrix}
1 & 0 \\
0 & e^{i\Delta\psi_{\mathrm{top}}}
\end{bmatrix}
\frac{1}{\sqrt{2}}  \begin{bmatrix}
1 & i\\
i & 1
\end{bmatrix}
\begin{bmatrix}
0 \\
\frac{1}{\sqrt{2}}
\end{bmatrix} \\
= 
& \frac{1}{2\sqrt{2}} \begin{bmatrix}
i(1+e^{i\Delta\psi_{\mathrm{top}}})\\
e^{i\Delta\psi_{\mathrm{top}}} - 1
\end{bmatrix} \\
\end{align*} 

Likewise, the transformation of the bottom arm is:
\begin{align*}
&M_{\mathrm{MMI}}
\begin{bmatrix}
1 & 0 \\
0 & e^{i\Delta\psi_{\mathrm{bottom}}}
\end{bmatrix}
M_{\mathrm{MMI}}
\begin{bmatrix}
\frac{1}{\sqrt{2}} \\
0
\end{bmatrix} \\
=
& \frac{1}{\sqrt{2}} \begin{bmatrix}
1 & i\\
i & 1
\end{bmatrix}
\begin{bmatrix}
1 & 0 \\
0 & e^{i\Delta\psi_{\mathrm{bottom}}}
\end{bmatrix}
\frac{1}{\sqrt{2}}  \begin{bmatrix}
1 & i\\
i & 1
\end{bmatrix}
\begin{bmatrix}
\frac{1}{\sqrt{2}} \\
0 
\end{bmatrix} \\
= 
& \frac{1}{2\sqrt{2}} \begin{bmatrix}
1 - e^{i\Delta\psi_{\mathrm{bottom}}})\\
i(1+e^{i\Delta\psi_{\mathrm{bottom}}})
\end{bmatrix} \\
\end{align*} 

We operate this device in the regime where $\sim$95\% of the light is directed to the tricoupler by the intensity modulators. For our conceptual understanding, we can approximate this as directing $\sim$all of the light to the bottom output of the second MMI in the top arm, and $\sim$all of the light to the top output of the second MMI in the bottom arm. Therefore, the intensity modulator TOPMs are set to voltages corresponding to a phase shift of \textpi for both $\psi_{\mathrm{top}}$ and $\psi_{\mathrm{bottom}}$. Consequently, the amplitude of the top arm input to the tricoupler is -1, and the amplitude of the bottom arm input to the tricoupler is 1, meaning that the two inputs already differ by a \textpi phase shift, conducive to nulling. This explains why the null voltage for the 1x5 tricoupler devices is close to 0 V, while the 1x3 tricoupler device achieves a null at a higher voltage around $\sim$12.5 V.

\subsubsection{Broadband nulling}
\label{sec:polychromatic}

\begin{figure}[t]
    \centering
    \includegraphics[width=0.5\linewidth]{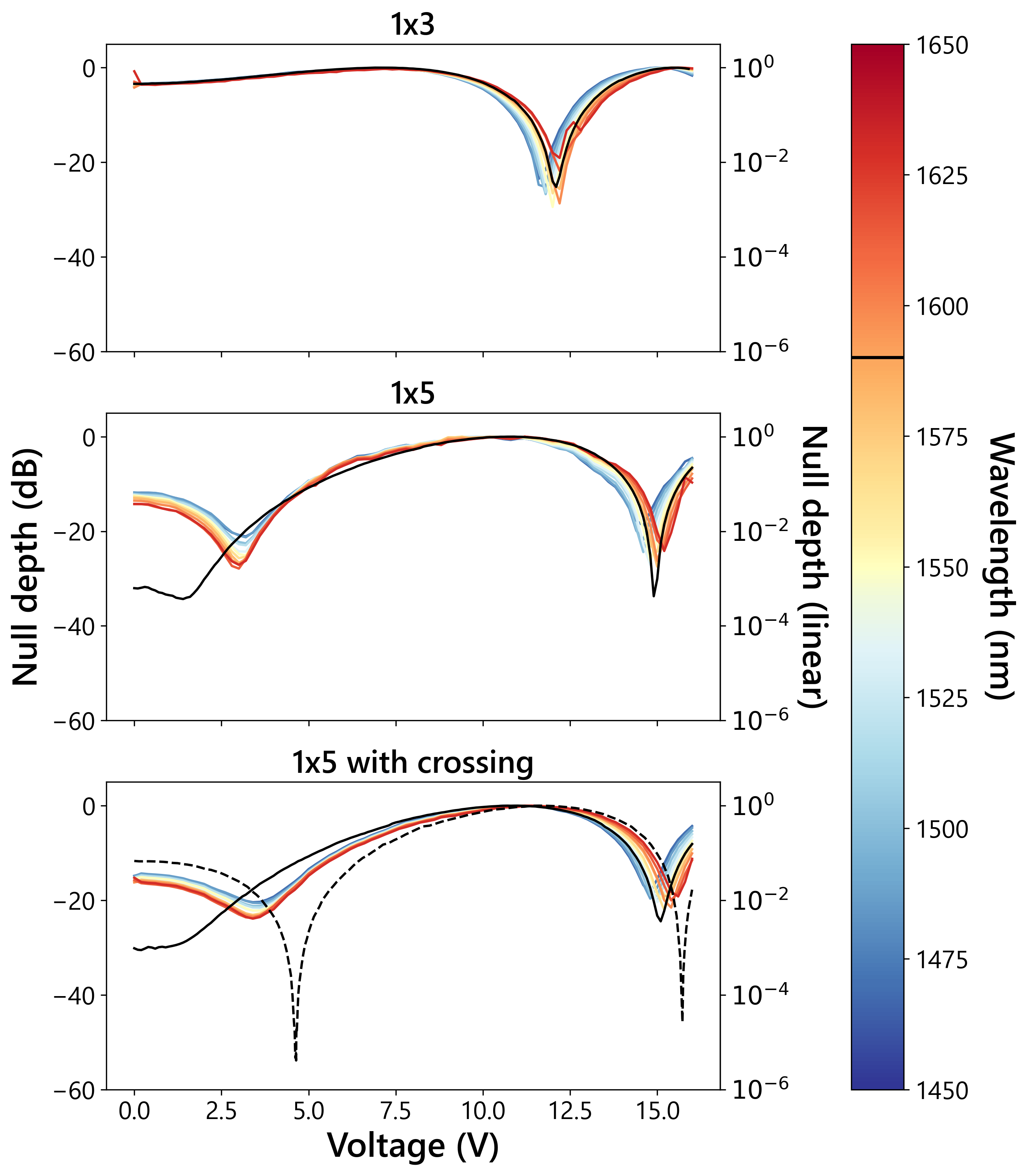}
    \caption{Null depths at different wavelengths (colored curves) compared to the monochromatic null depths obtained photometrically (black curves, dashed curve is polarization-filtered) for all tricoupler devices.}
    \label{fig:NullDepthPoly}
\end{figure}

To characterize the chromatic performance of the nulling circuits, we coupled a single-mode fiber to the nulled central port output of a given circuit and measured spectra with an OSA while sweeping the tricoupler input phase shifter. The polychromatic null depths of all 3 tricoupler devices (see Figure \ref{fig:NullDepthPoly}) present null depths consistent with the monochromatic null depths measured from imaging ($\sim$-30 dB). The colored curves in Figure \ref{fig:NullDepthPoly} reveal wavelength-dependent shifts in the phase shifter voltage required to achieve a null across all three tricoupler devices, indicating that the optimal voltage corresponding to a fixed optical phase is chromatic. Just as with the 1x2 MZI device in \S \ref{sec:MZI}, this is consistent with expectations, since applying a voltage heats the waveguide, inducing a phase delay by altering the effective index in a manner dependent on wavelength and temperature. This means that applying a single voltage to the phase shifter across the full bandpass will result in a degradation of the null away from the design wavelength by $\sim$1 order of magnitude over the bandpass. This behavior suggests that the observed broadband null depth is limited, in part, by the chromatic response of the TOPM.

In principle, the tricoupler null should be achromatic. In practice, to distinguish the intrinsic chromaticity of the tricoupler from that of the phase shifter, we compare with the wavelength-dependence of the null voltages of the 1x2 MZI from \S \ref{sec:MZI} fabricated with the same TOPM. By comparing the wavelength-dependent null voltages of the MZI and tricoupler devices, we can separate the heater-induced chromatic response from the residual wavelength dependence introduced by the tricoupler itself. This comparison provides a direct assessment of the intrinsic broadband nulling performance of the tricoupler architecture.

\begin{figure}[t]
    \centering
    \includegraphics[width=0.7\linewidth]{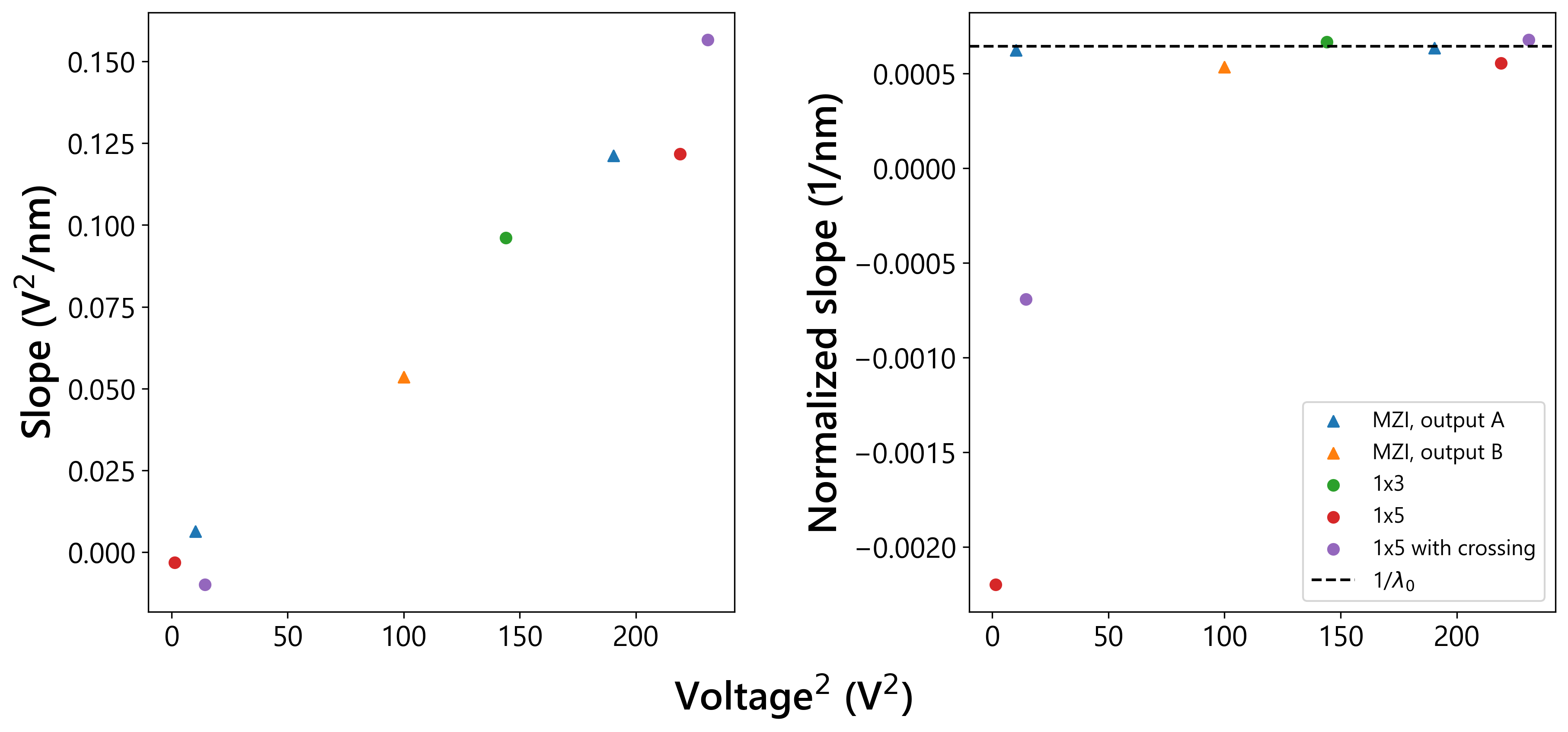}
    \caption{Chromaticity of the null voltage for MZI and tricoupler devices. Left: Measured slope of the null voltage squared, $dV^2/d\lambda$, as a function of the voltage squared at the null.  Right: The $V^2$-normalized slope, $(1/V^2)\,dV^2/d\lambda$, for the MZI and tricoupler devices. After normalizing by the heater power, the normalized slopes are close to constant, indicating that the dominant wavelength dependence originates from the thermo-optic phase shifter rather than the interferometric circuit.}
    \label{fig:NullChromaticity}
\end{figure}

The wavelength dependence of the null voltage is quantified by first fitting the device output modulation as a function and squared voltage, $V^2$, with a sinusoid at each wavelength. Since the thermo-optic phase shift is proportional to the dissipated heater power, and the heater power scales as $V^2$, the induced optical phase is expected to vary linearly with $V^2$. Next, we use the fitted modulation curves to determine the null voltages at each wavelength. Some devices have two nulls due to the $>2\pi$ phase stroke over the full 16 V voltage sweep. For all nulls of each device, we fit the slope of the null position in $V^2$ as a function of wavelength. As shown in Figure \ref{fig:NullChromaticity}, the raw chromatic slope increases with the heater power required to achieve the null, spanning more than an order of magnitude across the measured operating range. Because the thermo-optic phase shift is proportional to the dissipated heater power, which scales as $V^2$, higher-order nulls require larger heater powers and exhibit proportionally larger chromatic slopes. However, when the slope is normalized by the corresponding null voltage squared, the measurements collapse to a near-constant value. To understand why, consider a thermo-optic phase shift of the form:
\begin{equation}
\phi(V, \lambda) = \frac{2 \pi}{\lambda} L \Delta n(V, \lambda)
\end{equation}
where $V$ is the voltage applied to the heater, $\lambda$ is the wavelength of light, $L$ is the physical length of the heater, and $\Delta n(V, \lambda)$ is the index change induced by the heater
The heater-induced index change scales linearly with power dissipated, i.e. $\Delta n(V, \lambda) = \alpha V^2 f(\lambda)$, where $f(\lambda)$ captures any wavelength dependence of the heater's effect on refractive index, and so
\begin{equation}
    \phi(V, \lambda) = \frac{2 \pi}{\lambda} L\, \alpha\, V^2\,f(\lambda)
\end{equation}
Nulls occur at phases $\phi_\mathrm{null} = (2m+1)\pi$, where $m \in \mathbb{Z}$. Upon imposing this condition, we recover 
\begin{equation}
    V^2 = \frac{2m+1}{2\,L\,\alpha} \frac{\lambda}{f(\lambda)}
\end{equation}
Consequently
\begin{equation}
    \frac{dV^2}{d\lambda} = \frac{2m+1}{2\,L\,\alpha} \frac{f(\lambda) - \lambda f'(\lambda)}{f(\lambda)^2}
\end{equation}
Suppose the thermo-optic index change is dominated by heat dissipation and is relatively independent of wavelength as is expected for silica in the $H$ band \cite{Leviton2006}, i.e., $f(\lambda) \approx 1$. We can simplify the above expressions to 
\begin{align}
    V^2 &= \frac{2m+1}{2\,L\,\alpha} \lambda \\
    \frac{dV^2}{d\lambda} &= \frac{2m+1}{2\,L\,\alpha} \label{eq:NullChromaticity}\\
    \Rightarrow \frac{1}{V^2}\frac{dV^2}{d\lambda} &= \frac{1}{\lambda} \label{eq:NormalizedNullSlope}
\end{align}
Equation \ref{eq:NullChromaticity} implies successive nulls are increasingly sloped in chromaticity, as empirically demonstrated in the left panel of Figure \ref{fig:NullChromaticity}. However, it predicts the greatest relative change in slope between successive nulls should be between the first and second nulls, where the second null should be more sloped by a factor of 3 compared to the first null. Yet, as shown in the left panel of Figure~\ref{fig:NullChromaticity}, the slopes of successive MZI nulls (blue triangles), 1x5 tricoupler nulls (red circles), and 1x5 tricoupler nulls with a waveguide crossing (purple circles) differ by more than a factor of three. The underlying mechanism for the negative slope of the first nulls of the 1x5 tricoupler devices (red and purple points near 0 V in the left panel of Figure \ref{fig:NullChromaticity}) is not yet well understood and warrants further investigation.

Equation \ref{eq:NormalizedNullSlope} suggests the $V^2$-normalized chromatic slope of a null over a given bandpass should be constant at $1/\lambda_0$ if dominated by the chromaticity of the TOPMs, where $\lambda_0$ is the central wavelength of the bandpass. Indeed, this is consistent with normalized slopes for the MZI and tricoupler devices presented in the right panel of Figure \ref{fig:NullChromaticity}, with $1/\lambda_0$ at $\lambda_0 = 1550$ nm indicated by the black dashed line. This behavior indicates that the observed increase in chromaticity with heater power is primarily a consequence of the intrinsic wavelength dependence of the thermo-optic phase shifter, rather than an increase in the chromaticity of the interferometric circuit itself. Furthermore, the normalized slopes measured for the MZI and tricoupler devices are in good agreement with the dashed line, suggesting that both the tricoupler and the MZI introduce little additional wavelength-dependent phase beyond that expected from the common thermo-optic phase shifter.

\section{Discussion}
We demonstrate broadband nulling performance that highlights the potential of tricoupler-based photonic interferometers, including monochromatic null depths approaching -55 dB (4.1 $\times 10^{-6}$) and polychromatic null depths $\lesssim-30$ dB ($\lesssim 10^{-3}$) over a 200 nm bandpass. This is comparable to and in some cases exceeds the performance of recent two-telescope integrated optics beam combiners \cite{Sanny2026b, Martinod2021b}, which largely explore 3D beam combiners (directional couplers and tricouplers) fabricated by ultrafast laser inscription. Our results also reveal limitations that direct next steps for improved performance. The preliminary polarization-filtered nulls measured for the 1x5 tricoupler with crossing device indicates that polarization leaking arising from device birefringence predominantly limits the achievable extinction. Future implementations of these components may require two separate interferometric branches in the circuit to null orthogonal polarization states independently. Our empirical polychromatic nulls show that chromatic phase error introduced by the TOPMs limit the broadband nulling capabilities of both the MZI and tricoupler interferometric circuits, whereas intensity imbalances preliminarily do not appear to significantly compromise performance at the null depths obtained in this work; this remains to be confirmed by obtaining polarization-filtered null depths for the 1x3 tricoupler. Since the tricoupler nulls are inherently achromatic, broadband performance could be improved by developing achromatic phase shifters and intensity modulators. The devices we present do not include static phase delays, so the \textpi phase difference between the two interferometric arms for nulling is achieved by the combination of phase offsets applied by the MZI intensity modulators and the phase shifter on one of the tricoupler inputs. Static achromatic phase delays can be engineered by designing segmented waveguides of varying widths for the two interferometric arms upstream of the tricoupler to tailor the wavelength-dependent effective indices of each segment such that the differential phase difference accumulated between the two waveguides at the output is constant across the operating bandpass \cite{Klinner-Teo2022}. Realistically, the fabricated delays may deviate from the \textpi phase difference in practice, perhaps due to design errors, manufacturing defects, or fluctuating environmental conditions. To adjust for these practicalities, we are working on designing tunable achromatic phase shifters by adding TOPMs to different segments of the static variable-width phase delays. Likewise, we aim to develop spectrally-tunable intensity modulators to reduce wavelength-dependent intensity mismatch, preserving balanced interference over the full operating band. It also remains to be seen if contamination from scattered light coupling to the central nulled ports of our existing tricoupler devices substantially degrades the null depth.

Future device architectures could also target substantially deeper null depths with more sophisticated photonic circuits. Cascading multiple nulling stages would provide additional suppression of residual stellar leakage by rejecting light remaining after the initial null. More advanced architectures, such as kernel nullers, offer another promising direction by combining multiple interferometric outputs to produce observables that are intrinsically less sensitive to instrumental phase errors while enabling higher-order nulling \cite{Laugier2020}. The planar nature of lithographically fabricated integrated photonics necessitates waveguide crossings for compact routing, allowing these more complex beam-combination architectures to be implemented with a relatively small footprint while supporting multiplexed interferometric networks. As demonstrated in Figure~\ref{fig:NullDepthMonoAll}, waveguide crossings can be incorporated without significantly degrading tricoupler nulling performance.

Finally, the demonstrated single-input device characterization provides a foundation for our upcoming two-input interferometric operation. Injecting coherent light into both tricoupler inputs independently (as opposed to splitting on-chip as done in this work to simulate two coherent beams to interfere) would enable true stellar nulling while simultaneously exploiting the additional output ports for wavefront sensing. In particular, the bright ports and intensity modulator residual dump ports contain information about the amplitude and phase of the incoming wavefront that can be used for real-time sensing and control. This integrated approach could achieve simultaneous nulling, wavefront sensing, and science acquisition within a single photonic circuit, reducing non-common-path errors and providing a promising architecture for future high-contrast imaging instruments.

\section{Conclusions}

In this work, we demonstrated broadband nulling with integrated photonic tricoupler devices fabricated on a silica-on-silicon platform. By leveraging the symmetric interference properties of the tricoupler architecture, we seek to achieve achromatic nulling behavior across the measured bandpass around 1550 nm $\pm$ 100 nm, but are limited by the chromaticity of phase offsets induced by TOPMs. We demonstrate a monochromatic, polarization-filtered null depth of $\sim$54 dB (4.1 $\times$ 10$^{-6}$). These results highlight both the promise of integrated tricoupler architectures for compact broadband nulling interferometers and the importance of further optimization of device polarization performance and phase control elements. Future improvements, including achromatic phase and intensity modulation as well as more intricate nulling architectures offer pathways toward deeper and more robust nulls. The demonstrated compatibility of waveguide crossings with tricoupler nulling performance further enables the scaling of these devices toward more complex multiplexed interferometric networks for future high-contrast astronomical instrumentation.

\acknowledgments % equivalent to \section*{ACKNOWLEDGMENTS}       
 
This work was supported by the National Science Foundation under Grant No. 2308360. This work was supported by NASA under Grant No. 80NSSC24K1559. This research was carried out in part at the California Institute of Technology and the Jet Propulsion Laboratory under a contract with the National Aeronautics and Space Administration (NASA). The authors would also like to thank Michael Gutierrez for help upgrading the thermal control system to regulate the PIC's bulk temperature as well as Jake Zimmer for providing a pinhole mask and setting up a new power supply to improve measurement precision for the polarization-filtered experiment.

% References
\bibliography{references} % bibliography data in report.bib
\bibliographystyle{spiebib} % makes bibtex use spiebib.bst

\end{document}

%% file: commands.tex
\newcommand{\micron}{\textmu m\xspace}
\newcommand{\textlambda}{\ensuremath{\lambda}\xspace}
\newcommand{\textpi}{\ensuremath{\pi}\xspace}

%% file: references.bib
@ARTICLE{Fulton2021,
       author = {{Fulton}, Benjamin J. and {Rosenthal}, Lee J. and {Hirsch}, Lea A. and {Isaacson}, Howard and {Howard}, Andrew W. and {Dedrick}, Cayla M. and {Sherstyuk}, Ilya A. and {Blunt}, Sarah C. and {Petigura}, Erik A. and {Knutson}, Heather A. and {Behmard}, Aida and {Chontos}, Ashley and {Crepp}, Justin R. and {Crossfield}, Ian J.~M. and {Dalba}, Paul A. and {Fischer}, Debra A. and {Henry}, Gregory W. and {Kane}, Stephen R. and {Kosiarek}, Molly and {Marcy}, Geoffrey W. and {Rubenzahl}, Ryan A. and {Weiss}, Lauren M. and {Wright}, Jason T.},
        title = "{California Legacy Survey. II. Occurrence of Giant Planets beyond the Ice Line}",
      journal = {ApJS},
         year = 2021,
        month = jul,
       volume = {255},
       number = {1},
          eid = {14},
        pages = {14},
          doi = {10.3847/1538-4365/abfcc1},
archivePrefix = {arXiv},
       eprint = {2105.11584},
 primaryClass = {astro-ph.EP},
       adsurl = {https://ui.adsabs.harvard.edu/abs/2021ApJS..255...14F}
}

@ARTICLE{Morbidelli2000,
       author = {{Morbidelli}, A. and {Chambers}, J. and {Lunine}, J.~I. and {Petit}, J.~M. and {Robert}, F. and {Valsecchi}, G.~B. and {Cyr}, K.~E.},
        title = "{Source regions and time scales for the delivery of water to Earth}",
      journal = {MAPS},
         year = 2000,
        month = nov,
       volume = {35},
       number = {6},
        pages = {1309-1320},
          doi = {10.1111/j.1945-5100.2000.tb01518.x},
       adsurl = {https://ui.adsabs.harvard.edu/abs/2000M&PS...35.1309M}
}

@ARTICLE{Jovanovic2023,
       author = {{Jovanovic}, Nemanja and {Gatkine}, Pradip and {Anugu}, Narsireddy and {Amezcua-Correa}, Rodrigo and {Basu Thakur}, Ritoban and {Beichman}, Charles and {Bender}, Chad F. and {Berger}, Jean-Philippe and {Bigioli}, Azzurra and {Bland-Hawthorn}, Joss and {Bourdarot}, Guillaume and {Bradford}, Charles M. and {Broeke}, Ronald and {Bryant}, Julia and {Bundy}, Kevin and {Cheriton}, Ross and {Cvetojevic}, Nick and {Diab}, Momen and {Diddams}, Scott A. and {Dinkelaker}, Aline N. and {Duis}, Jeroen and {Eikenberry}, Stephen and {Ellis}, Simon and {Endo}, Akira and {Figer}, Donald F. and {Fitzgerald}, Michael P. and {Gris-Sanchez}, Itandehui and {Gross}, Simon and {Grossard}, Ludovic and {Guyon}, Olivier and {Haffert}, Sebastiaan Y. and {Halverson}, Samuel and {Harris}, Robert J. and {He}, Jinping and {Herr}, Tobias and {Hottinger}, Philipp and {Huby}, Elsa and {Ireland}, Michael and {Jenson-Clem}, Rebecca and {Jewell}, Jeffrey and {Jocou}, Laurent and {Kraus}, Stefan and {Labadie}, Lucas and {Lacour}, Sylvestre and {Laugier}, Romain and {{\L}awniczuk}, Katarzyna and {Lin}, Jonathan and {Leifer}, Stephanie and {Leon-Saval}, Sergio and {Martin}, Guillermo and {Martinache}, Frantz and {Martinod}, Marc-Antoine and {Mazin}, Benjamin A. and {Minardi}, Stefano and {Monnier}, John D. and {Moreira}, Reinan and {Mourard}, Denis and {Nayak}, Abani Shankar and {Norris}, Barnaby and {Obrzud}, Ewelina and {Perraut}, Karine and {Reynaud}, Fran{\c{c}}ois and {Sallum}, Steph and {Schiminovich}, David and {Schwab}, Christian and {Serbayn}, Eugene and {Soliman}, Sherif and {Stoll}, Andreas and {Tang}, Liang and {Tuthill}, Peter and {Vahala}, Kerry and {Vasisht}, Gautam and {Veilleux}, Sylvain and {Walter}, Alexander B. and {Wollack}, Edward J. and {Xin}, Yinzi and {Yang}, Zongyin and {Yerolatsitis}, Stephanos and {Zhang}, Yang and {Zou}, Chang-Ling},
        title = "{2023 Astrophotonics Roadmap: pathways to realizing multi-functional integrated astrophotonic instruments}",
      journal = {Journal of Physics: Photonics},
         year = 2023,
        month = oct,
       volume = {5},
       number = {4},
          eid = {042501},
        pages = {042501},
          doi = {10.1088/2515-7647/ace869},
archivePrefix = {arXiv},
       eprint = {2311.00615},
 primaryClass = {astro-ph.IM},
       adsurl = {https://ui.adsabs.harvard.edu/abs/2023JPhP....5d2501J}
}

@INPROCEEDINGS{Spalding2024,
       author = {{Spalding}, Eckhart and {Arcadi}, Elizabeth and {Douglass}, Glen and {Gross}, Simon and {Guyon}, Olivier and {Martinod}, Marc-Antoine and {Norris}, Barnaby and {Rossini-Bryson}, Stephanie and {Taras}, Adam and {Tuthill}, Peter and {Ahn}, Kyohoon and {Deo}, Vincent and {El Morsy}, Mona and {Lozi}, Julien and {Vievard}, Sebastien and {Withford}, Michael},
        title = "{The GLINT nulling interferometer: improving nulls for high-contrast imaging}",
    booktitle = {Optical and Infrared Interferometry and Imaging IX},
         year = 2024,
       editor = {{Kammerer}, Jens and {Sallum}, Stephanie and {Sanchez-Bermudez}, Joel},
       series = {Society of Photo-Optical Instrumentation Engineers (SPIE) Conference Series},
       volume = {13095},
        month = aug,
          eid = {1309507},
        pages = {1309507},
          doi = {10.1117/12.3016348},
       adsurl = {https://ui.adsabs.harvard.edu/abs/2024SPIE13095E..07S}
}

@ARTICLE{Laugier2023,
       author = {{Laugier}, Romain and {Defr{\`e}re}, Denis and {Woillez}, Julien and {Courtney-Barrer}, Benjamin and {Dannert}, Felix A. and {Matter}, Alexis and {Dandumont}, Colin and {Gross}, Simon and {Absil}, Olivier and {Bigioli}, Azzurra and {Garreau}, Germain and {Labadie}, Lucas and {Loicq}, J{\'e}r{\^o}me and {Martinod}, Marc-Antoine and {Mazzoli}, Alexandra and {Raskin}, Gert and {Sanny}, Ahmed},
        title = "{Asgard/NOTT: L-band nulling interferometry at the VLTI. I. Simulating the expected high-contrast performance}",
      journal = {A\&A},
         year = 2023,
        month = mar,
       volume = {671},
          eid = {A110},
        pages = {A110},
          doi = {10.1051/0004-6361/202244351},
archivePrefix = {arXiv},
       eprint = {2211.09548},
 primaryClass = {astro-ph.IM},
       adsurl = {https://ui.adsabs.harvard.edu/abs/2023A&A...671A.110L}
}

@ARTICLE{Sanny2026,
       author = {{Sanny}, Ahmed and {Labadie}, Lucas and {Gross}, Simon and {Barjot}, K{\'e}vin and {Laugier}, Romain and {Garreau}, Germain and {Martinod}, Marc-Antoine and {Defr{\`e}re}, Denis and {Withford}, Michael J.},
        title = "{Asgard/NOTT: L-band nulling interferometry at the VLTI: III. The mid-infrared integrated optics beam combiner for NOTT}",
      journal = {A\&A},
         year = 2026,
        month = jan,
       volume = {705},
          eid = {A37},
        pages = {A37},
          doi = {10.1051/0004-6361/202555865},
archivePrefix = {arXiv},
       eprint = {2511.19790},
 primaryClass = {astro-ph.IM},
       adsurl = {https://ui.adsabs.harvard.edu/abs/2026A&A...705A..37S}
}

@ARTICLE{Norris2020,
       author = {{Norris}, Barnaby R.~M. and {Cvetojevic}, Nick and {Lagadec}, Tiphaine and {Jovanovic}, Nemanja and {Gross}, Simon and {Arriola}, Alexander and {Gretzinger}, Thomas and {Martinod}, Marc-Antoine and {Guyon}, Olivier and {Lozi}, Julien and {Withford}, Michael J. and {Lawrence}, Jon S. and {Tuthill}, Peter},
        title = "{First on-sky demonstration of an integrated-photonic nulling interferometer: the GLINT instrument}",
      journal = {MNRAS},
         year = 2020,
        month = jan,
       volume = {491},
       number = {3},
        pages = {4180-4193},
          doi = {10.1093/mnras/stz3277},
archivePrefix = {arXiv},
       eprint = {1911.09808},
 primaryClass = {astro-ph.IM},
       adsurl = {https://ui.adsabs.harvard.edu/abs/2020MNRAS.491.4180N}
}

@ARTICLE{Martinod2021a,
       author = {{Martinod}, Marc-Antoine and {Tuthill}, Peter and {Gross}, Simon and {Norris}, Barnaby and {Sweeney}, David and {Withford}, Michael J.},
        title = "{Achromatic photonic tricouplers for application in nulling interferometry}",
      journal = {Appl. Opt.},
         year = 2021,
        month = jul,
       volume = {60},
       number = {19},
        pages = {D100},
          doi = {10.1364/AO.423541},
archivePrefix = {arXiv},
       eprint = {2106.00251},
 primaryClass = {astro-ph.IM},
       adsurl = {https://ui.adsabs.harvard.edu/abs/2021ApOpt..60D.100M}
}

@ARTICLE{Klinner-Teo2022,
       author = {{Klinner-Teo}, Teresa and {Martinod}, Marc-Antoine and {Tuthill}, Peter and {Gross}, Simon and {Norris}, Barnaby and {Leon-Saval}, Sergio},
        title = "{Achromatic design of a photonic tricoupler and phase shifter for broadband nulling interferometry}",
      journal = {Journal of Astronomical Telescopes, Instruments, and Systems},
         year = 2022,
        month = oct,
       volume = {8},
          eid = {045001},
        pages = {045001},
          doi = {10.1117/1.JATIS.8.4.045001},
archivePrefix = {arXiv},
       eprint = {2210.01040},
 primaryClass = {astro-ph.IM},
       adsurl = {https://ui.adsabs.harvard.edu/abs/2022JATIS...8d5001K}
}

@INPROCEEDINGS{Leviton2006,
       author = {{Leviton}, Douglas B. and {Frey}, Bradley J.},
        title = "{Temperature-dependent absolute refractive index measurements of synthetic fused silica}",
    booktitle = {Optomechanical Technologies for Astronomy},
         year = 2006,
       editor = {{Atad-Ettedgui}, Eli and {Antebi}, Joseph and {Lemke}, Dietrich},
       series = {Society of Photo-Optical Instrumentation Engineers (SPIE) Conference Series},
       volume = {6273},
        month = jun,
          eid = {62732K},
        pages = {62732K},
          doi = {10.1117/12.672853},
       adsurl = {https://ui.adsabs.harvard.edu/abs/2006SPIE.6273E..2KL}
}

@ARTICLE{Laugier2020,
       author = {{Laugier}, Romain and {Cvetojevic}, Nick and {Martinache}, Frantz},
        title = "{Kernel nullers for an arbitrary number of apertures}",
      journal = {A\&A},
         year = 2020,
        month = oct,
       volume = {642},
          eid = {A202},
        pages = {A202},
          doi = {10.1051/0004-6361/202038866},
archivePrefix = {arXiv},
       eprint = {2008.07920},
 primaryClass = {astro-ph.IM},
       adsurl = {https://ui.adsabs.harvard.edu/abs/2020A&A...642A.202L}
}

@ARTICLE{Hsiao2010,
       author = {{Hsiao}, Hsien-Kai and {Winick}, Kim A. and {Monnier}, John D.},
        title = "{Midinfrared broadband achromatic astronomical beam combiner for nulling interferometry}",
      journal = {Appl. Opt.},
         year = 2010,
        month = dec,
       volume = {49},
       number = {35},
        pages = {6675},
          doi = {10.1364/AO.49.006675},
archivePrefix = {arXiv},
       eprint = {1012.0790},
 primaryClass = {astro-ph.IM},
       adsurl = {https://ui.adsabs.harvard.edu/abs/2010ApOpt..49.6675H}
}

@ARTICLE{Martinod2021b,
       author = {{Martinod}, Marc-Antoine and {Norris}, Barnaby and {Tuthill}, Peter and {Lagadec}, Tiphaine and {Jovanovic}, Nemanja and {Cvetojevic}, Nick and {Gross}, Simon and {Arriola}, Alexander and {Gretzinger}, Thomas and {Withford}, Michael J. and {Guyon}, Olivier and {Lozi}, Julien and {Vievard}, S{\'e}bastien and {Deo}, Vincent and {Lawrence}, Jon S. and {Leon-Saval}, Sergio},
        title = "{Scalable photonic-based nulling interferometry with the dispersed multi-baseline GLINT instrument}",
      journal = {Nature Communications},
         year = 2021,
        month = jan,
       volume = {12},
          eid = {2465},
        pages = {2465},
          doi = {10.1038/s41467-021-22769-x},
       adsurl = {https://ui.adsabs.harvard.edu/abs/2021NatCo..12.2465M}
}

@inproceedings{Jovanovic2025,
author = {{Jovanovic}, Nemanja and {Kim}, Yoo Jung and {Sanny}, Ahmed and {Kenchington Goldsmith}, Harry-Dean and {Fitzgerald}, Michael P. and {Gatkine}, Pradip and {Mawet}, Dimitri},
title = {{Developing photonic components for nulling applications}},
volume = {13627},
booktitle = {Techniques and Instrumentation for Detection of Exoplanets XII},
editor = {Garreth J. Ruane and Maxwell A. Millar-Blanchaer},
organization = {International Society for Optics and Photonics},
publisher = {SPIE},
pages = {136270O},
year = {2025},
doi = {10.1117/12.3065787},
URL = {https://doi.org/10.1117/12.3065787}
}

@ARTICLE{Goldsmith2026,
       author = {{Goldsmith}, Harry-Dean Kenchington and {Jovanovic}, Nemanja and {Asnodkar}, Anusha Pai and {Kim}, Yoo Jung and {Sanny}, Ahmed and {Gatkine}, Pradip and {Fitzgerald}, Michael P.},
        title = "{Tri-coupler geometries for achromatic nulling interferometry in the near-infrared}",
      journal = {Optics Express},
         year = 2026,
        month = may,
       volume = {34},
       number = {10},
        pages = {19122},
          doi = {10.1364/OE.583815},
archivePrefix = {arXiv},
       eprint = {2602.23693},
 primaryClass = {astro-ph.IM},
       adsurl = {https://ui.adsabs.harvard.edu/abs/2026OExpr..3419122G}
}

@ARTICLE{Bracewell1978,
       author = {{Bracewell}, R.~N.},
        title = "{Detecting nonsolar planets by spinning infrared interferometer}",
      journal = {Nature},
         year = 1978,
        month = aug,
       volume = {274},
       number = {5673},
        pages = {780-781},
          doi = {10.1038/274780a0},
       adsurl = {https://ui.adsabs.harvard.edu/abs/1978Natur.274..780B}
}

@ARTICLE{Sanny2026b,
       author = {{Sanny}, Ahmed and {Gretzinger}, Thomas and {Gross}, Simon and {Labadie}, Lucas and {Withford}, Michael},
        title = "{Mid-infrared nulling interferometry beam combiners using asymmetric directional couplers}",
      journal = {Optics Letters},
         year = 2026,
        month = may,
       volume = {51},
       number = {10},
        pages = {2832},
          doi = {10.1364/OL.596119},
       adsurl = {https://ui.adsabs.harvard.edu/abs/2026OptL...51.2832S}
}
